\documentstyle[multicol,aps]{revtex}

\newcommand \bq {\begin{eqnarray}}
\newcommand \be {\begin{equation}}
\newcommand \eq {\end{eqnarray}}
\newcommand \ee {\end{equation}}

\begin{document}
\draft
\twocolumn
\title{Arrows of Time and the Anisotropic Properties of CMB}

\author{A.E. Allahverdyan$^{1,2)}$, V.G. Gurzadyan$^{2,3)}$
}
\address{$^{1)}$Institute of Theoretical Physics, University of Amsterdam
Valckenierstraat 65, 1018 XE Amsterdam, The Netherlands
\\ $^{2)}$Yerevan Physics Institute,
Alikhanian Brothers St. 2, Yerevan 375036, Armenia, \\
$^{3)}$ ICRA, Dipartimento di Fisica, Universita di
Roma La Sapienza, 00185 Rome, Italy
}

\maketitle

\begin{abstract}
The relation between the thermodynamical and cosmological arrows of time
is
usually viewed in the context of the initial conditions of the Universe.
It is a necessary but not sufficient
condition for ensuring the thermodynamical arrow. We point out that in the
Friedmann-Robertson-Walker Universe with negative curvature $k=-1$
there is the second necessary ingredient.
It is based on the geodesic mixing - the dynamical instability
of motion along null geodesics in hyperbolic space.
Kolmogorov (algorithmic) complexity as a universal and experimentally
measurable concept can be very useful in description of this chaotic
behavior using the data on Cosmic Microwave Background radiation.
The formulated {\it curvature anthropic principle} states the negative
curvature as a necessary condition for the time asymmetric Universe
with an observer.
\end{abstract}


\section{Introduction.}

The Universe is expanding thus determining the cosmological arrow of time.
The fate of the expansion, i.e. whether the Universe will expand forever
or
not, crucially depends on the mean density of the Universe.
The Universe is filled by Cosmic Microwave Background radiation (CMB) with
Planckian spectrum and highly isotropic temperature.

In the present paper we attempt to argue that {\it these observational
facts and the
thermodynamical arrow of time, and as the main consequence the second law
of
thermodynamics, both are the result of the geometry of the
Universe, namely, its negative curvature.} Thus the fate of the expansion
and
the second law are linked via the geometry of the Universe.

The association of the second law and the thermodynamical arrow with
the geometry of the Universe proposed below differs from other original
approaches on the nature of the thermodynamical arrow of
Gold \cite{Gold},
Penrose \cite{Penrose1} \cite{Penrose2},
Hawking \cite{Hawking} and Page \cite{Page},
Prigogine \cite{PP}, Zurek \cite{Zurek}, Zeh \cite{Zeh},
among the principal ones.

CMB is a cornerstone in our discussion. CMB photons are moving freely
during
the most
lifetime of the Universe, thus tracing its geometry. Obviously, CMB have
to
carry the signature of the geometry.
Indeed, the properties of CMB in hyperbolic Friedmann-Robertson-Walker
$(k=-1)$
Universe differ from those of flat, $k=0$, and positively curved, $k=+1$,
cases.
Particularly the exponential deviation of the geodesics and hence, the
effect of geodesic mixing in the hyperbolic spaces leads at least to the
following observable consequences \cite{g1}\cite{g2}:

(1) damping of anisotropy after the last scattering epoch;

(2) flattening of autocorrelation function;

(3) distortion of anisotropy spots.

It is remarkable, that the statistically significant signature of the
third
effect - the threshold independent distortion in the CMB sky maps -
has been detected by means of the analysis of COBE DMR 4-year data
\cite{g3}.
If it is really a result of the geodesic mixing, as predicted, then this
can be a direct indication of the negative curvature of the Universe, and
hence $\Omega_0 < 1$.

It is necessary to note, that at the present many observational facts -
particularly on the large-scale distribution of galaxies -
are supporting the mean density of the Universe to be below its critical
value. There is even an indication on the apparent accelerating rate
of the expansion, attributed to the contribution of the cosmological
constant
(e.g. \cite{Peeb}).

On the other hand, the time asymmetry of the Universe is one of the
deepest and yet challenging problems of modern physics.

Several different
and at first glance, independent arrows of time have been defined. \\
{\it (i)} The thermodynamical arrow of time - entropy of a 
closed system increases with time. \\
{\it (ii)} The cosmological arrow of time - the universe expands. \\
{\it (iii)} The psychological arrow of time - our knowledge about the past
is more definite than about the future.\\
{\it (iv)} The arrow of retarded electromagnetic interaction.

Although these arrows arise from CPT-symmetric dynamical laws, they
can be associated with CPT-asymmetric initial conditions as well;
below we deal with only T-invariance, since
CP-asymmetry of weak interactions is not directly relevant to the present
discussion \cite{Penrose1}.
The last two arrows are thought as closely related with the
thermodynamical arrow
\cite{davies}\cite{Penrose1}\cite{Penrose2}\cite{Zeh},
so that the main arrows are the cosmological and thermodynamical ones.

Thus in the present paper we claim that if the geometry
is basic for the cosmology, it may be
equally basic for the thermodynamics. First, we recall that
thermodynamical arrow for a closed system can be formulated as a
consequence of the following {\it necessary} conditions
\cite{zwanzig}\cite{Zeh}:

1) Absence of correlations in the initial conditions;

2) Dynamical chaos.

These two conditions appear already in the Boltzmann's derivation of his
famous kinetic equation, though perhaps not explicitly.
They can be traced out clearly in Zwanzig's derivation of
master-equation \cite{zwanzig} or Jaynes' information-theoretical
approach to irreversibility \cite{jaynes}. A usual discussion about
possible relations between cosmological and thermodynamical arrows of time
concentrates only the first condition \cite{Penrose1}-\cite{Page},
\cite{Zeh}, \cite{davies}.
Our purpose here is to point out that the
second ingredient of the thermodynamical arrow can have cosmological
context as well, and it is ensured due to the mixing of trajectories in
hyperbolic spaces.
Namely, if the Friedmann-Robertson-Walker universe has negative curvature,
then the flow of null geodesics which describes the free motion of
photons,
represents an Anosov system \cite{Anosov} - a class of dynamical systems
with
maximally strong statistical properties. Anosov systems are characterized
by exponential divergence of initially close trajectories, property of
K-mixing, positive KS-entropy, countable Lebesgue spectrum,
as well as by the exponential decay of the correlators \cite{Po}.

On the other hand, intrinsic decoupling of correlations is responsible
for the so-called Markovian behavior
\cite{arnold}\cite{zaslav}\cite{allah}.
As we shall discuss below this latter property can be one of two main
ingredients ensuring the thermodynamical
arrow \cite{shlogl}\cite{allah}, though the existence of other
mechanisms is not excluded.

Thus we show that the exponential instability of geodesic flow in
hyperbolic
FRW Universe - the geodesic mixing -  revealed through the properties of
CMB,
is relating the thermodynamical and cosmological arrows of time.

The paper is organized as follows. In section \ref{thermo} we
discuss the thermodynamical arrow and the
conditions which are necessary for its derivation; we follow mainly the
arguments of \cite{zwanzig} (see also \cite{Zeh})). In section
\ref{geo}
we show how the CMB mixing properties in the negatively curved space can
be
connected with the thermodynamical arrow.
Then we discuss the algorithmic complexity as a powerful tool for
description of the chaotic behavior under discussion (section IV).
Our conclusions are represented in the last section.

\section{Thermodynamical arrow of time}
\label{thermo}

Evolution of a closed non-relativistic system (the relativistic case will
add only technical difficulties)
is described by the Liouville equation for a probability density in the
phase space
\begin{equation}
\label{1}
{\rm i}\partial _t\rho =[H, \rho]\equiv {\cal L}\rho ,
\end{equation}
where ${\cal L}$ is the Liouville operator,
$H$ is the Hamiltonian, and $[..,..]$ denotes the Poisson bracket as
usual.
The quantum case can be investigated using exactly the same formulae, but
Poisson brackets should be substituted by commutators, and probability
densities by density matrices.
Thus, on the given question, neglecting certain effects \cite{alala},
 there are no principal differences between the classical and quantum
approach \cite{zwanzig}.
The statistical description is started by introducing a set of relevant
coarse-grained variables.
This means that the exact dynamical equation (\ref{1}) is projected into a
subspace of the phase space. Thus, a linear projection operator
$\Pi$ ($\Pi =\Pi ^2$)
is introduced, which transforms the density from the full phase space to
the subspace of the relevant variables
\begin{equation}
\label{2}
\Pi \rho = \rho _{rel},
\end{equation}
where $\rho _{rel}$ is the probability density of the relevant variables.
Dividing the initial equation (\ref{1}) into two parts which correspond
to $\rho _{rel}$ and $\rho _{ir}=\rho -\rho _{rel}$,
we get an equation for $\rho _{rel}$
\begin{eqnarray}
\label{3}
&& {\rm i} \partial _t \rho _{rel}=\Pi {\cal L}\rho _{rel} \nonumber \\
&& -{\rm i}\int _0^t G(\tau )\rho _{rel}(t-\tau ){\rm d}\tau +\Pi {\cal
L}\exp ({\rm i}(1-\Pi){\cal L}t)\rho _{ir},
\end{eqnarray}
\begin{equation}
\label{4}
G(t)=\Pi {\cal L}\exp (-{\rm i}(1-\Pi){\cal L}t)(1-\Pi ){\cal L}\Pi.
\end{equation}
Two essential assumptions must be imposed to get the thermodynamical arrow
of time in terms of the H-theorem.
First, the {\it autonomous} character of Eq. (\ref{3}) should be ensured:
$\rho _{rel}(t)$ can depend only on its initial conditions
$\rho _{rel}(0)$, and not on $\rho _{ir}(0)$.
It is necessary to assume that
\begin{equation}
\label{5}
\rho _{ir}(0)=0,
\end{equation}
in order to have {\it self-consistent} statistical description. We see
that this condition means the absence (in the proper sense) of initial
correlations in the system.
Conceptually this is very close to the initial conditions of the
Universe proposed by Penrose \cite{Penrose1}\cite{Penrose2}.
Ultimately, (\ref{5}) can be explained simultaneously via cosmological
initial conditions \cite{Zeh}.

However, even with the condition (\ref{5}) Eq. (\ref{3}) does not describe
the thermodynamical arrow or the second law.
Moreover, it is compatible with an anti-thermodynamical behavior.
The second necessary condition concerns the kernel $G(t)$ which arises
from the interaction of the relevant degrees of freedom with the
irrelevant
ones.
In other words, this kernel contains the contribution of the entire
history
of the relevant variables and hence, Eq. (\ref{3}) is non-Markovian.
However, it is by no means trivial that just this fact is incompatible
with the thermodynamic arrow \cite{shlogl}\cite{allah}.
We must have physical conditions supplying the Markovian (no-memory)
regime in Eq. (\ref{3}) \cite{shlogl}\cite{allah}.
Namely, let the kernel $G(t)$ varies much faster than $\rho
_{rel}(t)$; this allows us to consider $\rho _{rel}$
as constant during the characteristic times of variation of $G(t)$,
and hence, to
get Markovian, stationary master equation which is compatible with the
thermodynamical arrow
\begin{equation}
\label{6}
{\rm i} \partial _t \rho _{rel}=\Pi {\cal L}\rho _{rel}-{\rm i}\int
_0^{\infty}{\rm d}\tau G(\tau )\rho _{rel}(t).
\end{equation}
In the most general form the second law can be expressed by means of
decrease of {\it exergy} (this term was originally proposed by Clausius)
\begin{equation}
\label{6.5}
S(\rho (t))\leq S(\rho (t^{\prime})) \ \ {\rm iff} \ \ t\geq t^{\prime},
\end{equation}
where exergy (also called relative entropy or information gain) is an
useful generalization of the concept of thermodynamic potential
\begin{equation}
\label{7}
S(\rho (t))={\rm tr}\rho (\ln \rho -\ln \rho _{st}).
\end{equation}
Here tr means integration by the full phase space in the classical case,
and trace in the quantum one.
$\rho _{st}$ is a stationary state attained by the statistical system at
very large times. The latter, however, should be understood as
still much less than the Poincare recurrent time; the infinity at the
upper
limit in Eq. (\ref{6}) should be also considered in this way.
Therefore the recurrent time itself must be very large. This condition can
be considered as satisfied, since for majority of ``reasonable'' systems
the Poincare time exceeds the age of the Universe.
If this stationary distribution is the Gibbs' one,
we get free energy from Eq. (\ref{7}), and for the microcanonical
stationary state, describing the closed stat it is reduced to the {\it
minus} usual
Boltzmann-Gibbs-Shannon entropy.

It must be stressed again that the {\it both} above mentioned conditions
are necessary for the derivation of the
thermodynamical arrow. However, it is not the whole
story because the second assumption is essentially dynamical, and hence
needs
a concrete physical mechanism for its validity.
One of the possible mechanisms realizing the convergence to Markovian
behavior is the intrinsic chaoticity of the considered dynamical system
through the effective decoupling of correlators
\cite{mackay}\cite{zaslav}\cite{j}.
This property will be discussed in the next section.

At the end of this section let us mention that there are cases where
the markovian master equation (\ref{6}) can appear as too naive.
Namely, some additional, essentially system-dependent procedures
should be invoked to ensure the correct physical behavior.
However, they will not change the physical meaning of the markovian 
assumption.

\section{Geodesics mixing.}
\label{geo}

The geodesics of a space (locally if the space is non-compact) with
constant negative curvature in all two-dimensional directions
are known to possess properties of Anosov systems.

As it was proved by Pollicott \cite{Po} for dim=3 manifold $M$ with
constant negative curvature
the time correlation function of the geodesic flow $\{f^\lambda\}$
on the unit tangent bundle $SM$ of $M$
$$
b_{A_1,A_2}({\lambda})=\int_{SM}A_1\circ f^{\lambda}\cdot A_2 d\mu
-\int_{SM}A_1d\mu\int_{SM}A_2d\mu
$$
is decreasing by exponential law for all functions $A_1,A_2\in L^2(SM)$
\begin{equation}
   \left|b_{A_1,A_2}(\lambda)\right|
      \leq c\cdot \left|b_{A_1,A_2}(0)\right|\cdot e^{-h\lambda} \ ,
\label{expmix}
\end{equation}
where $c>0$, $\mu$ is the Liouville measure and $\mu(SM)=1$,
$h$ is the KS-entropy of the geodesic flow $\{f^{\lambda}\}$.

To reveal the properties of the free motion of photons in
pseudo-Riemannian (3+1)-space the projection of its geodesics
into Riemannian 3-space has to be performed, i.e. by corresponding
a geodesic $c(\lambda)=x(\lambda)$ to the geodesic in the former space:
$\gamma(\lambda)=(x(\lambda),t(\lambda))$.

Then the transformation of the affine parameter is as follows \cite{LMP}.
$$
  \lambda(t)=\int_{t_0}^t \frac{ds}{a(s)} \ .
$$
The KS-entropy in the exponential index can be easily estimated
for the matter dominated post-scattering Universe \cite{g1},
so that
\begin{equation}
   e^{h\lambda}=
      (1+z)^2\left[\frac{1
+\sqrt{1-\Omega}}{\sqrt{1+z\Omega}+\sqrt{1-\Omega}}\right]^4 \ .
\label{factor}
\end{equation}
i.e. depends on the density parameter $\Omega$ and the redshift of the
last scattering epoch $z$.
The initial condition (\ref{5}) should be
the one to ensure the thermodynamical arrow.

The decay of correlators for geodesic flow for $k=-1$ FRW Universe
provides the procedure of coarse-graining and ensures the
Markovian (no-memory) behavior of the CMB parameters (\ref{6}), so that
\begin{equation}
\label{c1}
t\gg \tau = 1/h,
\end{equation}
where KS-entropy defines the characteristic time scale $\tau$ and depends
only
on the diameter of the Universe (the only scale in the maximally symmetric
space) \cite{g1}.
The time scale $\tau$ is the so-called Markov time or a random variable
independent on future as defined in the theory of Markov processes
\cite{Dyn},
and in CMB problem describes the decay of initial perturbations,
i.e. damping of the initial anisotropy amplitude and the flattening of the
angular correlation function \cite{g1}, \cite{g2}.
For certain dynamical systems that time scale defines also
the relaxation time for tending to a microcanonical equilibrium.
The evaluation of $\tau$ and hence of the negative curvature of the
Universe
has been performed in \cite{g3} using the COBE data.

The negative constant curvature leads to a decay of time correlators
of geodesics, thus defining the thermodynamic arrow for CMB in FRW $k=-1$
Universe.

\section{Complexity as a Measure of Chaotic Behavior}

For the description of the chaoticity as  an underlying mechanism ensuring
the Markovian behavior, which is one of the two necessary ingredients of
the
thermodynamical arrow of time, the algorithmic complexity can appear to be
a convenient descriptor.

Algorithmic complexity was introduced  by
Kolmogorov \cite{kolmogorov} and independently by Solomonoff and Chaitin
\cite{chaitin}.
Mathematical aspects of the application of complexity to dynamical systems
are discussed in \cite{alekseev}.
The notion of algorithmic complexity has been discussed
extensively also for physical problems \cite{zurek}\cite{caves1} as well.
Algorithmic complexity (or algorithmic information) for an object
represented by a sequence of bits is defined as the length in bits of
the shortest possible program according to which
an universal Turing machine (computer),
starting from its standard and fixed initial state, produces
this object as its single output and halts. The universality of the
machine
ensures almost machine-independence of the considered definition.
As shown the machine-dependent part of
the shortest program can be neglected for relatively long sequences
\cite{kolmogorov}\cite{chaitin}.

Thus, the definition of the algorithmic complexity for an object $A$ reads
$$
K_A={\rm min}_{p} |p(A)|,
$$
where the minimization is made over all programs $p$ computing the object
for a fixed universal computer, and $|...|$ denotes the length in
bits as usual.

It is natural to call an object complex (or random) if its complexity is
comparable with
its length \cite{alekseev}. Hence, a chaotic system will possess bigger
complexity
than a regular one.
We are interested in {\it specific complexity}
\cite{alekseev}
$$
k(A)=\frac{K_A}{|A|},
$$
so that a finite non-zero $k(A)$ will mean that the object $A$ is
algorithmically complex.
As shown 
random sequences are indistinguishable (for all practical purposes) from
the ones generated by the proper stochastic process \cite{zvonkin}.
Notice that the time of computation does not enter in the definition of
complexity; this is the first hint that the exact calculation of
complexity can be hopeless. Moreover, it was proved
\cite{zvonkin}
that complexity is not a computable quantity. In fact, it is necessary for
its self-consistent definition.
In order to define complexity for a dynamical system one should consider a
translation of its trajectory
into a symbolic language \cite{alekseev}\cite{arnold}.
Namely, the phase space is divided
into non-overlapping regions, and every such a region gets its symbol.
Then a trajectory of the considered dynamical system can be viewed as a
sequence of symbols (merely as a matter of convenience
this sequence can be translated into the language of bits). Now the
dynamics
can be called
chaotic (for fixed initial conditions) if the corresponding
symbolic sequence is algorithmically complex
\cite{alekseev}\cite{ford}. Note that the above mentioned partition should
be detailed enough because algorithmic complexity is well-defined only for
sufficiently long sequences of symbols.

Besides the natural form of this definition, it has well-defined
advantages compared to the more traditional one through Lyapunov
exponents.
Indeed, the traditional signatures of chaos are less obvious if the
attention
is shifted from the
time evolution of the phase-space points to the time evolution of
probability
densities, governed by the coarse-grained Liouville equation. Namely, if
the distance between
two densities is defined in terms of an overlap integral, there is no
exponential divergence of initially close densities \cite{caves1}.
On the other hand, needless to say that trajectories and small shifting
of initial conditions cannot be available in many systems of
practical interest. It is the main reason of a recent activity seeking
unambiguous relations between KS-entropy and directly measurable
quantities like excess of entropy \cite{dzh}.
On the other hand, the definition through algorithmic complexity is
already given in the terms of coarse-grained description which is
reasonable
also in the general context of the thermodynamical arrow of time.

The fundamental theorem of Brudno and
Alekseev \cite{alekseev} states, that if KS-entropy is non-zero for a
dynamical
system, then there is the specific algorithmic complexity.
It is also natural that in a random sequence there are no long-range
correlations.  Such a correlation will introduce some ``order'',
and lead to a decrease of complexity. Thus, the definition through
algorithmic complexity allows to illustrate effectively this point also.

In the case of the CMB problem, the convenience of use of the complexity
is
determined by two reasons. First, one can derive a relation between the
curvature of the Universe and the complexity of the CMB anisotropy spots
\cite{g}, and, second, a concrete algorithm can be created enabling to
calculate the complexity of the spots at the CMB sky maps
\cite{ags}. The algorithm developed in \cite{ags} showed that the
complexity
is indeed a calculable quantity for digitized CMB data, and its values are
well
correlating with the values of the fractal dimension of the spots, and
therefore the complexity can be a powerful descriptor for the
analysis of the data of forthcoming CMB experiments.  Other
algorithms, particularly based on the random sequence concepts, may be
developed as well.

Thus the complexity allows to illustrate effectively the chaos-negative
curvature link through the CMB properties.

\section{Conclusion}

Thus, the negative curvature of the FRW Universe and the effect of
geodesic mixing can provide the condition
necessary for the emergence of the thermodynamic arrow of time.
Moreover this mechanism can explain why CMB  contains the
major fraction of the entropy of the Universe \cite{s}.

If this is indeed the mechanism of the origin of the thermodynamic arrow,
then the thermodynamics in a flat and positively curved universes not
necessarily to be strongly time asymmetric, and the latter is observed
since
we happened to live in a Universe with negative curvature.
In other words, the symmetry of the Newtonian mechanics,
electrodynamics, quantum mechanics might purely survive in some universes.
On the other hand, a recent activity devoted to the foundations of
thermodynamics \cite{lieb}
allows to disentangle time-asymmetric elements from the remained basis.
We are going to consider separately the relations of the above-mentioned
theory with the possible
thermodynamics description arising in the flat or negatively curved
space \cite{alala}.

In this context the essence of thermodynamical arrow must be understood as
not the mere
increase of entropy of an almost closed system, but the fact that this
arrow
has the universal direction in the entire Universe (see \cite{GMH}).
In the light of our suggested explanation
of the emergence of this arrow, it follows that the negative curvature is
the very mechanism unifying all local thermodynamical arrows. While
in the flat or positively curved universes, i.e. at the absence of a 
global unification mechanism, there can be local thermodynamical arrows
with various directions.

Another intriguing problem arising here, is whether life can occur in such
globally time symmetric universes,
or the time asymmetry/negative curvature is a necessary ingredient for
developing of life - {\it curvature anthropic principle}.
CMB features have to carry the signature of this principle.

We thank Th.M. Nieuwenhuizen and W. Zurek for valuable discussions.

\end{document}